\documentclass[preprint2,twoside]{hwo}

\usepackage{graphicx}
\usepackage{tabularx}

\input{hwo.h}

\begin{document}

\title{\textbf{\LARGE Debris disks and their properties with the Habitable Worlds Observatory}}
\author {\textbf{\large Isabel Rebollido$^{1}$ }, Yasuhiro Hasegawa$^{2}$, Meredith MacGregor$^{3}$, Bin Ren$^{4}$, Mark Booth$^{5}$, Jonathan Marshall$^{6}$, Courtney Dressing$^{7}$, Patricia Luppe$^{8}$ }
\affil{$^1$\small\it European Space Agency (ESA), European Space Astronomy Centre (ESAC), Camino Bajo del Castillo s/n, 28692 Villanueva de la Cañada, Madrid, Spain}
\affil{$^2$\small\it  Jet Propulsion Laboratory, California Institute of Technology, Pasadena, CA 91109, USA}
\affil{$^3$\small\it  Department of Physics \& Astronomy, John Hopkins University, 3400 North Charles Street, Baltimore, MD 21218, USA}
\affil{$^4$\small\it  Observatoire de la Cˆote d’Azur, 96 Bd de l’Observatoire, 06304 Nice, France}
\affil{$^5$\small\it  UK Astronomy Technology Centre, Royal Observatory Edinburgh, Blackford Hill, Edinburgh
EH9 3HJ, UK}
\affil{$^6$\small\it  Institute of Astronomy and Astrophysics, Academia Sinica, 11F of AS/NTU Astronomy-Mathematics Building, No.1, Section 4, Roosevelt Rd, Taipei 106319, Taiwan}
\affil{$^7$\small\it DepartmentofAstronomy,UniversityofCaliforniaBerkeley,Berkeley,CA94720,USA}
\affil{$^7$\small\it DepartmentofAstronomy,UniversityofCaliforniaBerkeley,Berkeley,CA94720,USA}
\affil{$^7$\small\it School of Physics, Trinity College Dublin, College Green, Dublin 2, Ireland}
\author{\footnotesize{\bf Endorsed by: Narsireddy Anugu (Georgia State University), 
Nicholas Ballering (Space Science Institute), 
Aarynn Carter (STScI), 
Gianni Cataldi (National Astronomical Observatory of Japan), 
Miguel Chavez Dagostino (National Institute of Astrophysics, Optics and Electronics), 
Denis Defrère (KU Leuven), 
Vincent Esposito (Chapman University), 
Ryan Fortenberry (University of Mississippi), 
Luca Fossati (Space Research Institute, Austrian Academy of Sciences), 
Eunjeong Lee (EisKosmos (CROASAEN), Inc.), 
Briley Lewis (University of California Santa Barbara), 
Briley Lewis (UCSB), 
Meredith MacGregor (Johns Hopkins University), 
Stanimir Metchev (University of Western Ontario), 
Patricio Reller (University College London), 
Pablo Santos-Sanz (Instituto de Astrofísica de Andalucía (CSIC), Spain.), 
Antranik Sefilian (University of Arizona), 
Sarah Steiger (Space Telescope Science Institute), 
Schuyler Wolff (University of Arizona)}.}



\begin{abstract}
  {The study of the last stages of planet formation, also known as debris disks, is fundamental to place constrains on the formation of planetary sized bodies. Debris disks are composed of dust and occasionally small amounts of gas, both released through dynamical interactions of small rocky bodies and dust particles, such as collisions and evaporation. The distribution of the dust can reveal the presence of forming planets and its composition can directly trace that of comets, asteroids and even planets. While we have been observing debris disks for 40 years now, most observations so far have been restricted to the cold outer regions of the system, and therefore information of the terrestrial zone is still missing. The improved spatial resolution, inner working angle and sensitivity that the Habitable Worlds Observatory will provide will enable a much closer look into the structure and composition of debris disks (particularly of its inner region) and enable the search for the forming rocky planets within the disk.}
  \\
  \\
\end{abstract}

\vspace{2cm}

\section{Science Goal}

Exploration of the environment of planetary systems is one of the next steps to understand planet formation and search for life. Debris disks are a direct way to investigate the composition, architecture and dynamical activity of young planetary systems, where terrestrial planet formation might still be ongoing.

Debris disks represent the last stage of planet formation, after the protoplanetary material is cleared through photoevaporation and accretion processes. They are sustained through interaction of planetesimals and dust particles that collide among each other to produce finer dust and then aggregate into larger particles, in a cycle that sustains a significant amount of dust of different sizes. This dust is detectable through various mechanisms, mainly through the light scattered by the particles and the thermal emission caused by the heat received from the central star.
The importance of debris disks resides in the fact that, while giant planets most likely form in the protoplanetary phase, rocky planet formation is most likely still ongoing, and so are dynamical interactions between planets, protoplanets and minor bodies. Through the study of the dust composition and its spatial and size distributions, we can infer characteristics of the whole system, such as the locations of planets in the disk, the composition of rocky planets and whether the interaction between dust, small bodies and planets can affect the atmospheres of the planets in the system, which is fundamental for habitability studies. The Habitable Worlds Observatory (HWO) will be uniquely able to resolve the dust at different wavelengths, revealing structures in debris disks and potentially unveiling planetary companions embedded within them.

\section{Science Objective}
\label{sec:sci_obj}
HWO will have the capability to look at the finer structures in debris disks, allowing the investigation of their composition, origin and role in shaping planetary systems. 

\subsection{The composition and origin of dust in debris disks}
\begin{figure*}
    \centering
    \includegraphics[]{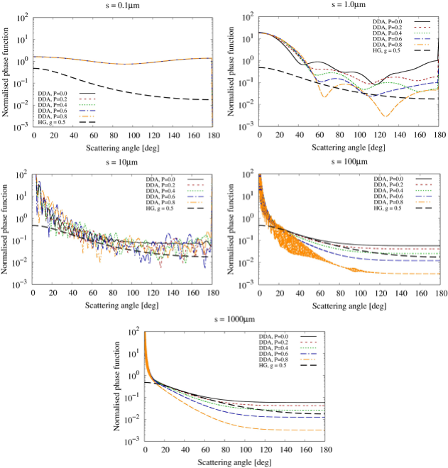}
    \caption{Figure from \citep{Pawellek24}. The scattering phase function varies for different sizes and dust compositions}
    \label{fig:SPF}
\end{figure*}

Multiwavelength total intensity and polarimetric observations: what is the dust composition?
Dust in debris disks is expected to be generated through collisions, mostly between cm to mm dust particles that, when destroyed, generate smaller particles that collide among themselves, and so on, in a process called collisional cascade. 
This is currently the dominant hypothesis, but it is challenging to address observationally, as the interpretation of observations rely on models that have degeneracies between the size and composition of the dust. The investigation of the scattering phase function (SPF, \ref{fig:SPF}), which is a proxy of how the dust reflects the stellar light, is one of the most used approaches for inferring the dust composition. SPF is challenging to measure, and often polarimetric observations are required (e.g. Millar-Blanchaer et al. 2016). Pawellek et al. (2024) discuss how different models (Mie, DDA, H-G) can be more or less effective at reproducing the observations, and the effects of gas in the dust distribution. While these studies are performed in the brightest debris disks (LIR/L* >10-5), higher contrasts are required for fainter disks, expected to be around more mature stars, where planetary systems are already formed.
The gain in sensitivity and spatial resolution from HWO will be crucial to reach less bright and more compact disks, providing insights into Kuiper Belt-like structures.  

Additionally, other techniques complementary to high contrast imaging, like polarimetry, can provide additional insights into the dust size distribution in the disk (Graham et al. 2007).

\subsection{Reveal fine debris disk structures to understand the dynamics of planetary systems. }

High contrast imaging with coronagraphy at high spatial resolution to investigate the distribution of the dust.
Disk morphology has been used to predict the presence of planetary mass companions (Pearce et al. 2022) and understand the dynamics of the system. 
The location of dust in debris disks can trace the dynamics in the system, like resonances with planets, collisions and even self induced processes like self stirring or P-R drag.
Our ability to trace these structures is sometimes determined by the geometry of the disk itself (i.e. inclination), but recent deprojection attempts (Gáspár et al. 2023) have shown that with high spatial resolution it is possible to gain a better understanding of the dust location. 
Recent findings (Marino et al., in prep., ARKS project) show that the relative distribution of small (observed in scattered light) vs large (observed in millimetre) dust suggests there's often (but not always) an offset between these components.
The breakthrough in terms of sensitivity that JWST/MIRI represents has revealed that collisions might be a common occurrence in debris disks (Rebollido et al. 2024, Fig. 2). However, due to the limitations of inner working angle (IWA) and spatial resolution in the mid-IR, we still don’t fully understand the distribution of dust in the inner regions of debris disks, including the terrestrial zone (i.e. exozodis). A new leap in sensitivity and smaller IWA in the near-IR with state of the art coronagraphy and improved data reduction techniques (e.g. stray stellar light suppression) can enable for the first time the investigation of the terrestrial zone of debris disks and therefore, the dynamics of dust in this inner regions of planetary systems. 

\begin{figure*}
    \centering
    \includegraphics[]{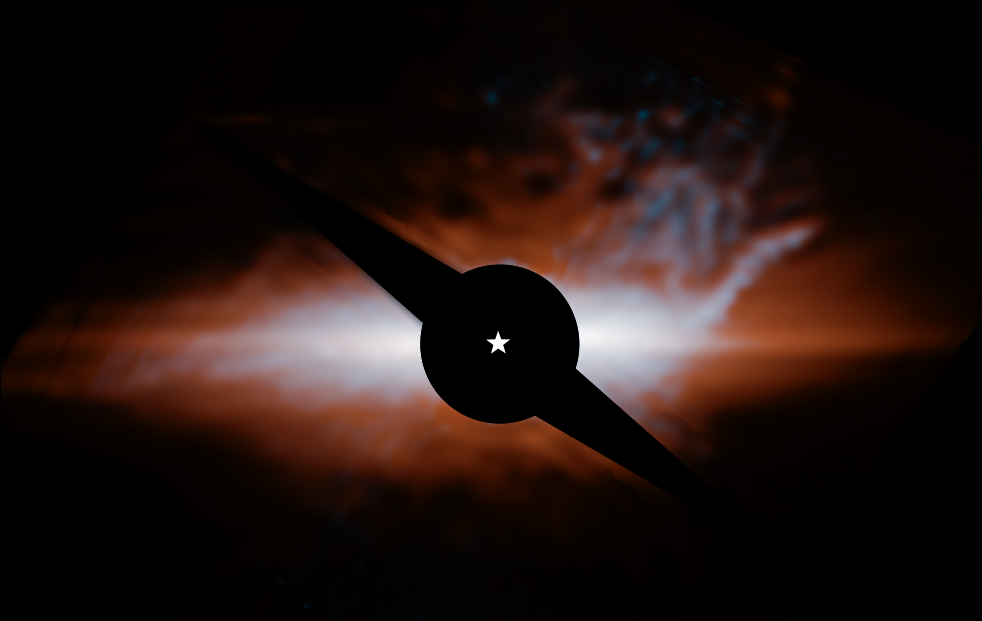}
    \caption{Figure from \cite{Rebollido24}. The debris disk around Beta Pictoris seen with JWST/MIRI.}
    \label{fig:placeholder}
\end{figure*}

\subsection{Dynamics in debris disks: is there a volatile delivery mechanism? }

{\textit Integral Field Spectroscopy of disks: where is the gas, can we observe water?}
After the discovery of debris disks 40 years ago, they were defined as gas free, in contrast with gas rich protoplanetary disks. Gas was expected to either accrete onto the star or the recently formed giant planets, or depleted from the system through radiative pressure. High resolution spectroscopy in visible and UV wavelengths revealed that there were small amounts of refractory elements in gaseous state, and the increase in sensitivity in sub-mm wavelengths thanks to ALMA has revealed in the past years significant amounts of volatile gas (mainly CO). This has changed the view of debris disks as gas free environments, and has raised the question of the origin of the gas. 
The most prevalent hypothesis right now is that collisions between bodies and dynamical interactions throwing material close to the inner regions of the systems might be releasing gas through evaporation and sublimation processes. These bodies that we can trace in the inner regions are also known as exocomets or, classically, falling evaporating bodies (FEBs). 
Particularly in the case of Beta Pictoris, where gas has been observed in multiple wavelengths and is consistent with a large number of bodies sublimating close to the star, FEBs have been linked to the gravitational interactions between the planets and the disk. This would mean that the presence of gas in debris disks could point to the presence of planets. Moreover, the release of volatile gas trapped in icy bodies from outer regions of planetary systems might be polluting the atmospheres and surfaces of rocky planets in the terrestrial zone, in a similar scenario to water delivery has been hypothesized for the Earth. So far, water has not yet been detected in a gaseous state in debris disks. 
Integral field spectroscopy with high resolution (R>80000) at near-UV (1400-1600 Å) and near-IR (1.5-3 µm) on board HWO will be crucial for understanding volatile delivery mechanisms in exoplanetary systems and enable the detection of water.

\section{Physical Parameters}
\label{sec:phys_par}
\subsection{Composition of the dust}
In order to significantly improve the understanding of dust composition and size distribution, high precision polarimetric observations will be needed, aiming at polarization levels below 1\%. Multi wavelength imaging and low resolution integral field spectroscopy that can reach small (<0.05 arcsec) resolution elements will also allow for investigation of the dependance of the composition with the location of dust in the disk.  

\subsection{Substructures and asymmetries}
The increase in spatial resolution in the past years thanks to high contrast imaging instruments like SPHERE, Gravity, or NIRCam and MIRI on board of the JWST have revealed unseen structures up to spatial resolutions of ~0.05 arcsec. Additional improvements in mirror size could push down to 0.03 arcsec, which would mean we could look at 0.1 au resolution for the closest disks, enough to resolve substructures analogous to the Asteroid belt’s Kirkwood gaps, induced by dynamical interaction with Jupiter. It is yet currently unknown whether the distribution of small grains can be explained by radiation pressure alone, and if so, do we get a constraint on stellar wind strength from these observations?
\subsection{Gas in debris disks}
While significant advances were made in the past years thanks to sub-mm observations, this wavelength can only trace the gas in the outer regions, which is observed in emission. In order to investigate warm and hot gas, located in the terrestrial zone, which might be released by evaporating and sublimating bodies, we require high sensitivity (SNR> 1000 in near-IR and near-UV), high resolution (>80000) spectroscopy.

\section{Description of Observations}

The unparalleled improvements from HWO will enable observations of debris disks in unprecedented detail, allowing for the characterization of the dust composition and distribution, the investigation of the gaseous component, and indirectly the search for planets in these systems.

\subsection{Observational Requirements}

Below are listed some of the requirements that would be needed for HWO to achieve the science goals listed above.

\begin{table*}[ht]
\begin{tabularx}{\textwidth}{lllll}
\hline
Capability & Details &  Wavelength  & Example references & Science objective \\
\hline
UV observations   & \begin{tabular}[c]{@{}l@{}}Spectroscopic FUV\\ R$\gtrsim$80000\end{tabular} &  1400-1600 Å  & \begin{tabular}[c]{@{}l@{}}\cite{Roberge2000}\\ \cite{Brennan24}\end{tabular} & \begin{tabular}[c]{@{}l@{}}Composition of rocky bodies \\through sublimation products. \\Detection of organics.\end{tabular}\\
\hline
\begin{tabular}[c]{@{}l@{}}NIR observations \\(\textgreater 1.5 microns)\end{tabular} & \begin{tabular}[c]{@{}l@{}}Composition of the dust \\Structures in the disk\\ Polarimetry \\ R$\gtrsim$80000 \end{tabular}&  1 - 3 µm & \begin{tabular}[c]{@{}l@{}}\cite{Ren23}\\ \cite{Olofsson19} \end{tabular}& \begin{tabular}[c]{@{}l@{}}NIR is necessary to investigate the \\distribution of fine dust particles \\in the disk and their composition. \\Additionally, spectroscopy could \\reveal the presence of water (2.8 µm) \\and/or CO (4.5 µm)\end{tabular} \\
\hline
High spatial res. & \begin{tabular}[c]{@{}l@{}}Resolution elements of\\ $\sim$0.03 arcsec \end{tabular} & $\sim$1 µm & \begin{tabular}[c]{@{}l@{}}\cite{Gaspar23}\\ \cite{Rebollido24} \end{tabular}&  \begin{tabular}[c]{@{}l@{}}Structures in debris disks are small \\(\textless{}1 au). Detection of planets and \\planet/disk interactions. \end{tabular} \\
\hline
High spectral res.  &  R\textgreater{}80000 & 1400 Å - 3 µm & \begin{tabular}[c]{@{}l@{}}\cite{Roberge2000}\\ \cite{Rebollido20} \end{tabular} & \begin{tabular}[c]{@{}l@{}}Gas features in debris disks are faint \\and with small velocity ranges \\(\textless{}50-100 km/s), so high spectral \\resolution is needed to detect them. \end{tabular}\\
\hline
Field of view  & \begin{tabular}[c]{@{}l@{}} 2’x2’ (approx. size of \\Fomalhaut)\end{tabular} & 1400 Å - 3 µm & \cite{Gaspar23} &  \begin{tabular}[c]{@{}l@{}}While spatial resolution is key to\\ investigate dust distribution, a field of \\view containing the whole disk will \\make observations more efficient.\end{tabular}\\                                                
\end{tabularx}
\label{tab:requirements}
\end{table*}

The requirements listed on Table \ref{tab:requirements} respond directly to the science objectives and physical parameters from Sect. \ref{sec:sci_obj} and \ref{sec:phys_par}. 
In order to resolve the structure of debris disks, we estimate that a resolution of at least 0.03 arcsec will be necessary, as it corresponds to 1 au at roughly 30 pc (for reference, the distance to Beta Pic is $\sim$ 20 pc). At the same time, the field of view (FOV), needs to cover at least 2'x2' in order to enable studies of the largest disks \citep[e.g. Fomalhaut][]{Gaspar23}.

In order to investigate the gas, observations through the whole wavelength coverage should enable at least R$\sim$80000, allowing for the detection of the circumstellar gas features, and avoiding blending with ISM or stellar components in the spectra. 

Finally, spectropolarimetric observations will enable the investigation of solid state features that reveal the composition of the dust, but only if the resolution is large enough to distinguish different spectral features (e.g. silicates; R>3000) and the polarimeter is able to reach polarization fractions well below the 10\% expected for debris disks \citep{Hull22}.

\subsection{Why HWO Is Needed}

The unique combination of HWO's space-based, diffraction-limited imaging and high-resolution spectroscopy in the UV-optical makes it the only facility capable of delivering this science:
Debris disks are challenging targets that require high spectral and spatial resolution, along with a wide wavelength range coverage. Ground based observatories face the unavoidable challenge of atmospheric contamination from near to far infrared, thus rendering almost impossible to reach the cold material in debris disks. Additionally, if we want to investigate the dust distribution in debris disks, it is fundamental that a high spatial resolution is delivered. While larger telescopes are possible on the ground, the stability that a space mission provides is key to reach resolution elements below 0.05 arcsec. 
The Habitable Worlds Observatory will be the only facility able to deliver the requested science.

\acknowledgements
Y.H. was supported by the Jet Propulsion Laboratory, California Institute of Technology, under a contract with the National Aeronautics and Space Administration (80NM0018D0004).

\bibliography{author.bib}

\end{document}